\documentclass[conference]{IEEEtran}
\IEEEoverridecommandlockouts
\usepackage{cite}
\usepackage{amsmath,amssymb,amsfonts}
\usepackage{physics}
\usepackage{algorithmic}
\usepackage{textcomp}
\usepackage{xcolor}
\usepackage{flushend}
\def\BibTeX{{\rm B\kern-.05em{\sc i\kern-.025em b}\kern-.08em
    T\kern-.1667em\lower.7ex\hbox{E}\kern-.125emX}}
\ifCLASSINFOpdf
  \usepackage[pdftex]{graphicx}
  \graphicspath{{/}}
  \DeclareGraphicsExtensions{.eps}
\else
  \usepackage[dvips]{graphicx}
  \graphicspath{{/}}
  \DeclareGraphicsExtensions{.pdf,.eps,.png}
\fi

\usepackage{babel,blindtext}
\begin{document}

%\title{Understanding Magnetic and Magneto-Quasistatic Human Body Communication in Comparison with Electro-Quasistatic HBC

\title{Understanding The Role of Magnetic and Magneto-Quasistatic Fields in Human Body Communication}

\author{Mayukh~Nath,
        Alfred~Krister~Ulvog, Scott~Weigand,
        and~Shreyas~Sen,~\IEEEmembership{Senior~Member,~IEEE}% 
\thanks{M. Nath, A. Ulvog and S. Sen are with the School of Electrical and Computer Engineering, Purdue University, West Lafayette, IN 47907, USA (e-mail: nathm@purdue.edu).

S. Weigand is with Eli Lilly and Company, Indianapolis, IN 46221, USA.}}

\maketitle

\begin{abstract}
With the advent of wearable technologies, Human Body Communication (HBC) has emerged as a physically secure and power-efficient alternative to the otherwise ubiquitous Wireless Body Area Network (WBAN). Whereas the most investigated nodes of HBC have been Electric and  Electro-quasistatic (EQS) Capacitive and Galvanic, recently Magnetic HBC (M-HBC) has been proposed as a viable alternative. Previous works have investigated M-HBC through an application point of view, without developing a fundamental working principle for the same. In this paper, for the first time, a ground up analysis has been performed to study the possible effects and contributions of the human body channel in M-HBC over a broad frequency range (1kHz to 10 GHz), by detailed electromagnetic simulations and supporting experiments. The results show that while M-HBC can be successfully operated as a body area network, the human body itself plays a minimal or negligible role in it’s functionality. For frequencies less than $\sim$30 MHz, in the domain of operation of Magneto-quasistatic (MQS) HBC, the human body is transparent to the quasistatic magnetic field. Conversely for higher frequencies, the conductive nature of human tissues end up attenuating Magnetic HBC fields due to Eddy currents induced in body tissues, eliminating the possibility of the body to support efficient waveguide modes. With this better understanding at hand, different modes of operations of MQS HBC have been outlined for both high impedance capacitive and 50$\Omega$ termination cases, and their performances have been compared with EQS HBC for similar sized devices, over varying distance between TX and RX. The resulting report presents the first fundamental understanding towards M-HBC operation and its contrast with EQS HBC, aiding HBC device designers to make educated design decisions, depending on mode of applications.
\end{abstract}


\section{Introduction}

\begin{figure}
\centering
     \includegraphics[width=0.87\columnwidth]{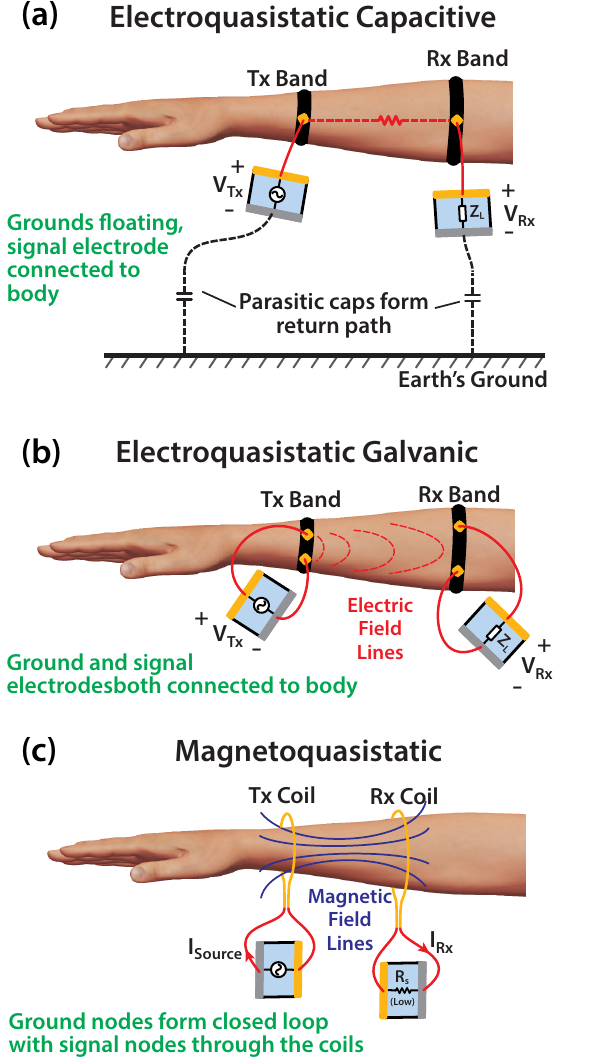}
     \caption{Different quasistatic HBC modalities depending on the placement of signal and ground planes: (a) Electro-Quasistatic Capacitive, (b) Electro-Quasistatic Galvanic, and (c) Magneto-Quasistatic HBC.}
     \label{fig:basics}
\end{figure}

In the last decade, Internet-of-Things has become the new buzzword in the community, and with that, the importance of wearable technology has increased many-fold. As more and more wearable devices start to appear in different fields such as personal assistant,  healthcare, secure authentication and so on, the need to develop a secure and efficient communication modality is becoming more and more apparent. In recent years, Human Body Communication (HBC) \cite{zimmerman_personal_1996, callejon_2013,wegmueller_channel,das_nsr_2019,wegmueller_signal_2010,park_embc_2015,Lucev_2011,Ruiz_2005} has emerged as a viable alternative to Wireless Body Area Network (WBAN), which is the present day standard of communication between wearable devices. HBC uses the human body as a transmission medium for signal \cite{zimmerman_personal_1996}. 
\begin{figure*}
    \centering
    \includegraphics[width=\textwidth]{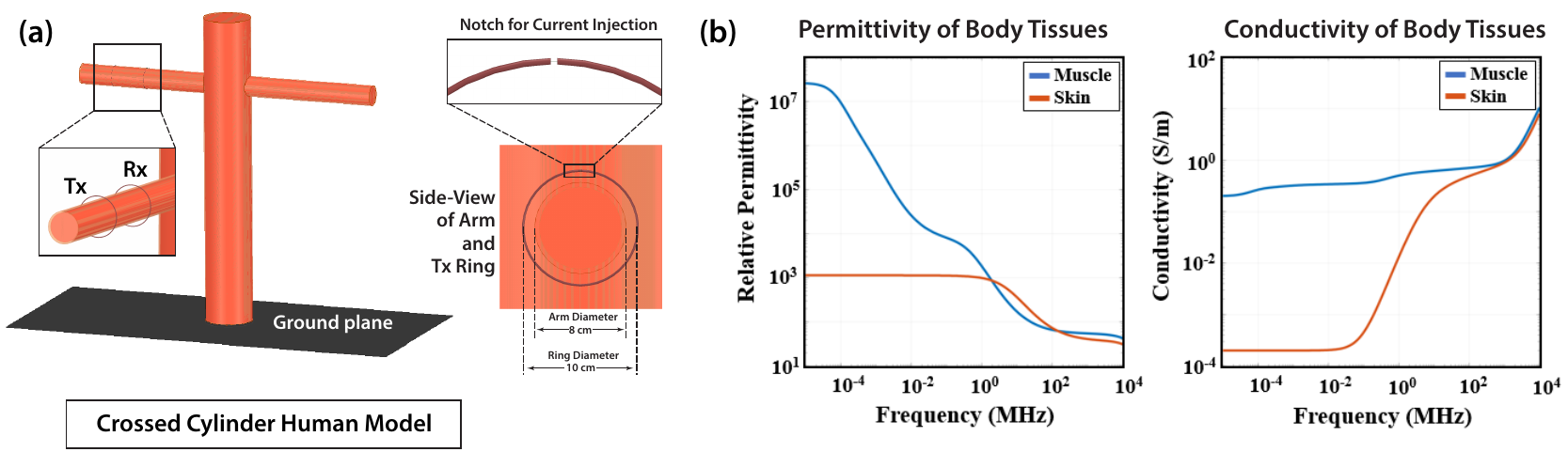}
    \caption{(a) Simplified crossed cylindrical model used for HFSS simulations. Two circular rings are used as transmitter and receiver. This simplified model was cross-checked with respect to a detailed complex human model\cite{neva_model} to ensure validity.  (b)  Relative permittivity ($\epsilon_r$ and conductivity ($\sigma$) of muscle and skin, used in HFSS simulations. Data was taken from the works of Gabriel\cite{gabriel_measurements}}
    \label{fig:sim_setup}
\end{figure*}
HBC devices can be divided into \textit{Electric} and \textit{Magnetic}, depending on whether the signal is coupled to the body using electrodes, or current coils respectively. Further, a particular HBC protocol can be labeled \textit{Quasistatic} when the frequency of operation is low enough - so that the corresponding wavelength is large compared to dimensions of the human body \cite{das_nsr_2019}. Fig. \ref{fig:basics} shows the different modalities of quasistatic HBC, depending on how the signal and ground planes interact with the human body or the environment. First, in \textit{Electro-Quasistatic (EQS) Capacitive HBC}, the signal planes are coupled to the body using electrodes, while the ground planes are left floating. The floating ground planes couple through parasitic capacitances with earth's ground to form the return path in capacitive HBC. Second, for \textit{EQS Galvanic HBC}, both signal and ground planes are coupled to the body using electrodes. At the TX, signal and ground nodes form electric field lines - part of which is picked up differentially at the receiver. Finally, in \textit{Magneto-Quasistatic (MQS) HBC}, the signal and ground planes are connected through a coil. Current flowing through the coil create alternating magnetic field. Part of this magnetic flux goes through the receiver coil, and induces an EMF that is picked up by the receiver circuitry.

The human body channel in EQS HBC had been studied in detail,\cite{maity_biophysical_2018,wegmueller_channel,callejon_2013,gomez_I2MTC_2017,maity_wearable_2017,Lucev_2011, nath_return_path_2020} especially for the case of capacitive HBC. On the other hand, very few works  have been done to understand MQS HBC, or in fact Magnetic HBC (M-HBC) in general. \cite{park_embc_2015,gomez_I2MTC_2017} presented a technique for M-HBC that utilizes a capacitor to cancel out the inductance of the TX or RX coil, to improve channel response. \cite{park_isscc_2019} has utilized the technique developed by \cite{park_embc_2015} to build a M-HBC transceiver to stream audio to in-ear headphones. These works have primarily focused on applications, performing minimal theoretical analysis of the Human Body Channel for Magnetic HBC. We also note that \cite{park_embc_2015} and \cite{park_isscc_2019} have claimed that the human body can function as a waveguide to carry M-HBC signals through the body. However, as we will show, that statement may not necessarily be true. In this paper, for the first time, we perform a detailed analysis through simulation and experiments, to fundamentally understand the strengths and limitations of the Magnetic Human Body Channel. Specifically, we seek the answer to the following questions:
\begin{itemize}
    \item How does the human body channel behave in the context of Magnetic Human Body Communication (M-HBC), and fundamentally where does that behaviour originate?
    \item Knowing the human body channel characteristics for M-HBC, what are the practical considerations for executing MQS HBC, and how does the response compare with it's EQS HBC counterparts?
\end{itemize}

The rest of the paper is organized as follows$:$ in Sec. II, we introduce an idealized excitation model and a simple human body model - used for the purpose of EM simulation; Sec. III consists of detailed analysis of body channel in magnetic HBC using EM simulations; Sec. IV follows with practical MQS HBC setups and measurement results, and finally the paper concludes in Sec. V.

\begin{figure*}
    \centering
    \includegraphics[width=1\textwidth]{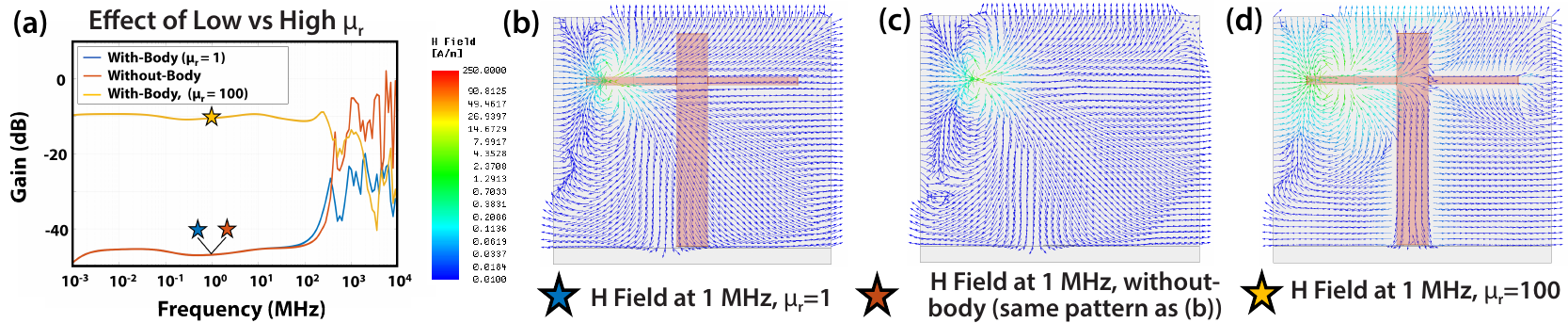}
    \caption{(a) Channel gain comparison between simple human body models with regular and artificially high $\mu_r$. High $\mu_r$ model increases the magnetic coupling, and hence channel gain. Compared to the without-body case, gain for the regular $\mu_r$ case is the same in low frequency, and lower in high frequency (f$>$100MHz). Vector H field plots at 1MHz for (b) $\mu_r$ = 1 case, (c) without-body case and (d) Artificially high $\mu_r$ case. Note that (b) and (c) are identical, denoting that human body does not affect the magnetic fields.}
    \label{fig:mu_plot}
\end{figure*}

\section{Idealized Excitation Model for Magnetic HBC}

\label{sec:excitation_model}

As we have introduced in Fig. \ref{fig:basics}, loops made of conducting wire are fundamental transmitting and receiving elements in Magnetic HBC. In a real-world device, the transfer characteristics between any two devices are affected by the channel characteristics, as well as output and input impedance of the transmitter and the receiver respectively. To de-embed the effects of termination impedance and to investigate the effect of human body channel for magnetic HBC, an idealized excitation model is designed for the purpose of electromagnetic field simulations. Simulations are performed in High Frequency Structure Simulator (HFSS), an FEM based Maxwell's equations solver from ANSYS. The transmitting ring, as shown in Fig \ref{fig:sim_setup}(a), is designed with a notch, and an ideal current source is placed, to inject a sinusoidal current into the ring. The receiving ring is a complete loop to allow free flow of current, and the induced current in this ring is measured and compared with the constant transmitter current to evaluate channel response. Since magnetic field is fundamentally generated from a current in the quasistatic regime, use of an ideal current source eliminates any effect of termination impedance, aiding a simpler analysis.

Further, we also create a simplified human body model using two crossed cylinders, as shown in Fig. \ref{fig:sim_setup}(a). The structure has a 4mm thick outer layer that has dielectric properties of dry human skin, while the interior has properties of averaged human muscle. Simulations using this simple structure provide intuitions about different fundamental aspects of the Magnetic HBC body channel in a fast and reliable manner. Although this simplified simulation model may appear as an oversimplification, a previous work by Maity et al\cite{hbc_safety} had explored electric and magnetic field distributions in and around the human body for HBC through extensive EM simulations. They compared field distributions between a similar simplified crossed cylinder model, and a complex human model – VHP Female v2.2 from Neva Electromagnetics\cite{neva_model} to be precise – and shown that the field distributions inside and outside the model, at least in the skin and muscle tissues, look identical. Since most human body tissues have really high relative dielectric permittivity, the simplification including only skin and muscle can be made without loosing much generality - while at the same time aiding us to perform a range of simulations within a reasonable amount of time and computational complexity. The relative permittivity ($\epsilon_r$) and conductivity ($\sigma$) of skin and muscle tissues in the simulation model have been taken from the works of Gabriel et al\cite{gabriel_measurements}, and shown here in Fig. \ref{fig:sim_setup}(b). In some simulations, these properties are intentionally altered for improved understanding, as will be described in Sec. \ref{sec:hfss_analysis}.

The transmitting and receiving rings have been kept on the same arm of the crossed cylinder model, and the distance between them is maintained at a constant 20 cm for the analysis in Sec. \ref{sec:hfss_analysis}. The common radii of the rings is chosen to be 5cm and the thickness of the wire 2mm, making the rings slightly larger compared to the arms - which have a radius of 4cm. Finally, the rings and the arm are made concentric, ensuring absence of direct contact between skin and the rings - this eliminates the possibility of current in the rings shorting through conductive body tissues.

\section{Significance of the Human Body Channel in Magnetic HBC}
\label{sec:hfss_analysis}

In this section, we explore contribution of the human body channel in Magnetic HBC, by EM simulations in HFSS using the simple model and ideal excitation described in the previous section. We perform frequency sweeps in the range of 1 kHz to 10 GHz and explore significance (or lack thereof) of three key parameters, namely relative permeability ($\mu_r$), permittivity ($\epsilon_r$), and conductivity ($\sigma$).

\subsection{Effect of relative permeability ($\mu_r$)}

If the transmitter and receiver rings are imagined as the primary and secondary coils of a transformer, presence of a magnetic core with $\mu_r>1$ would improve the transfer ratio between the rings. Unfortunately though, $\mu_r$ is 1 for human body tissues, and so the human body is transparent to static and quasistatic magnetic fields. If there is any improvement in signal transfer at all due to the presence of a human body, that cannot be explained through relative permeability.

It will however, still be interesting to artificially increase the $\mu_r$ of the tissue materials in the simulation model, to see how it affects the signal transfer. Because first and foremost, an increase in coupled signal due to increased $\mu_r$ would provide validation for the simulation setup. But also, there can be special medical application cases such as bone grafting, bionic limbs etc - where it is possible to manipulate $\mu_r$ - enabling magnetic HBC implementation of monitoring/ communicating devices for these cases. Keeping that in mind, we perform a simulation with an artificially high $\mu_r$ of 100 of the body model. Figure \ref{fig:mu_plot}(a) compares the transmission characteristics between $\mu_r=1$, $\mu_r=100$, and the baseline case (where body is absent). For $f<100$ MHz, we see a huge rise in gain for the high $\mu_r$ case, as expected, where as the $\mu_r=1$ case is largely identical with the without-body case. We do note that for $f>100$ MHz, the response from the without-body case is higher compared to with-body for both $\mu_r =$ 1 and 100. This behaviour cannot be explained just yet; we will be dealing with this in the following sections.

\subsection{Effect of relative permittivity ($\epsilon_r$)}

\begin{figure*}
    \centering
    \includegraphics[width=0.9\textwidth]{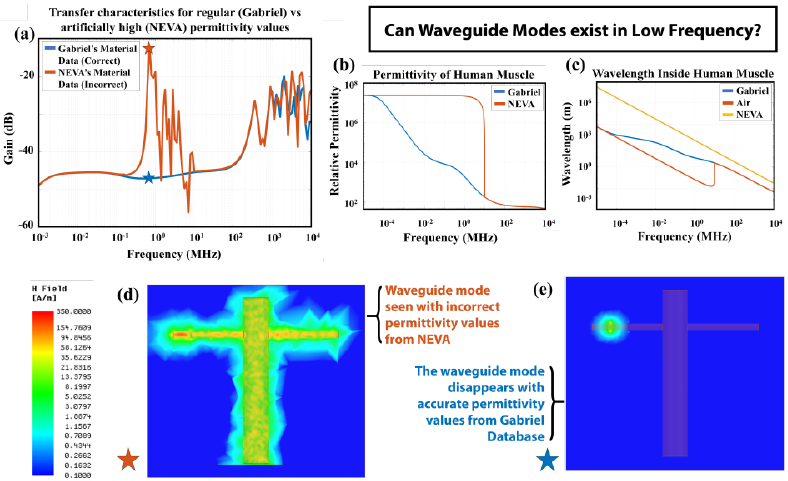}
    \caption{(a) Comparison of channel gain with accurate $\epsilon_r$ from Gabriel database vs Incorrect NEVA database (accurate above 10 MHz). Unusually high $\epsilon_r$ from NEVA model results $\lambda$ comparable to body dimensions in (100kHz-10MHz) and creates waveguide modes. (b) Deviation of the incorrect $\epsilon_r$ of NEVA data (linearly interpolated between 10 Hz and 10 MHz), from Gabriel (c) $\lambda$ calculated based on (b), showing that $\lambda$ gets unusually low for the NEVA data. (d) H field plot for simulation with NEVA data shows strong waveguide mode at 700 kHz. (e) Waveguide mode disappears, when accurate $\epsilon_r$ from Gabriel database is used.}
    \label{fig:waveguide_low_freq}
\end{figure*}

Unlike the case of $\mu_r$, the relative permittivity $\epsilon_r$ is high compared to the air ($\epsilon_{r,air}\approx1$). As shown in Fig. \ref{fig:sim_setup}(b), $\epsilon_r$ of the human tissues are orders of magnitude higher than unity for $f<1$ MHz, and it's value drops with increasing frequency. As wavelength $\lambda\propto1/\sqrt{\epsilon_r}$, higher $\epsilon_r$ of the body implies a lower $\lambda$ compared to air, i.e. $\lambda_{Body}\ll\lambda_{air}$. The question that follows, is whether such a high discontinuity in $\lambda$ can give rise to confinement of electro-magnetic waves in the body - i.e. whether the body can support waveguide modes.

% \begin{equation}
% \lambda = \frac{v}{f} = \frac{1}{f \sqrt{\epsilon\mu}} = \frac{c}{f\sqrt{\epsilon_r\mu_r}}
% \label{eqn:lambda}
% \end{equation}

As a side-track, an interesting direction towards answering this question came from using an incomplete tissue property database for our simulations. Initially, we borrowed material properties for skin and muscle from the detailed HFSS human model from NEVA Electromagnetics - the one we had mentioned in Sec. II. Using these properties, we obtained a result shown in Fig. \ref{fig:waveguide_low_freq}(a) - where in the frequency range 100 kHz - 10 MHz, the human body seemed to significantly enhance Magnetic HBC transfer. In fact, the H field plot at the peak at $f=707$ kHz, shown in \ref{fig:waveguide_low_freq}(d), resembled a waveguide mode. What we had initially failed to notice however, is that NEVA's database included accurate dielectric properties for only $f>10$ MHz. For $f<10$ MHz, it was simply a linear interpolation between 10 Hz and 10 MHz. This resulted into an unusually high $\epsilon_r$ (and hence unusually low $\lambda$) compared to the accurate Gabriel database \cite{gabriel_measurements}. When plotted in log-scale axes (Fig. \ref{fig:waveguide_low_freq}(b) and (c), it is apparent that the deviation from the accurate model is maximum just before 10 MHz; and for the 100 kHz - 10 MHz range, the $\lambda$ of the body becomes comparable or less than the human-body dimensions (the diameter of the arm in this case, which is 8 cm). This made waveguide modes possible in the simulated body structure, resulting the high signal transfer. The message obtained from this exercise then, is that while the low frequency waveguide modes were inaccurate, \textit{the high discontinuity between $\lambda$ between the human body and air should indeed be able to support wave-guide modes, as long as $\lambda$ is comparable or smaller than body dimensions.} 

\begin{figure}[!b]
    \centering
    \includegraphics[width=0.9\columnwidth]{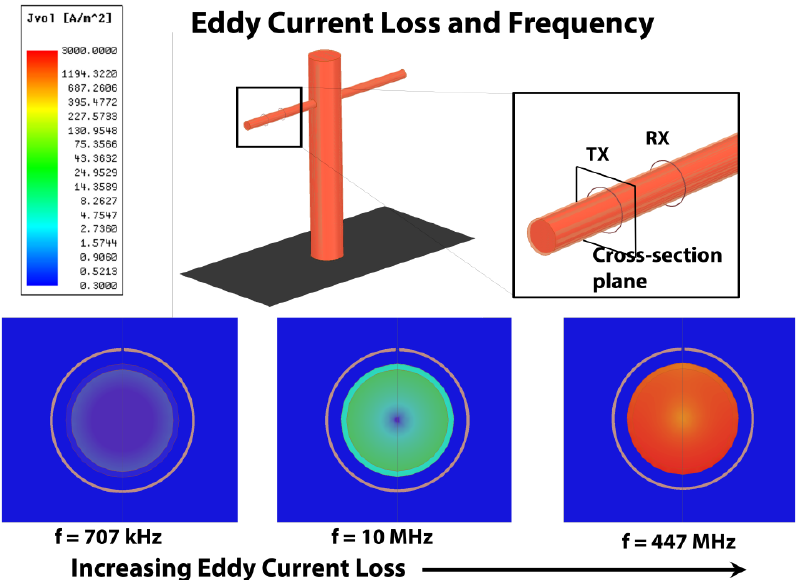}
    \caption{Current density (J) plot of the cross section of the model’s arm at the position of TX ring. As conductivity increases in higher frequency according to Fig. \ref{fig:sim_setup}(b), J also increases - signifying increased loss due to Eddy currents at higher frequency. }
    \label{fig:eddy_current_plot}
\end{figure}
\begin{figure*}
    \centering
    \includegraphics[width=\textwidth]{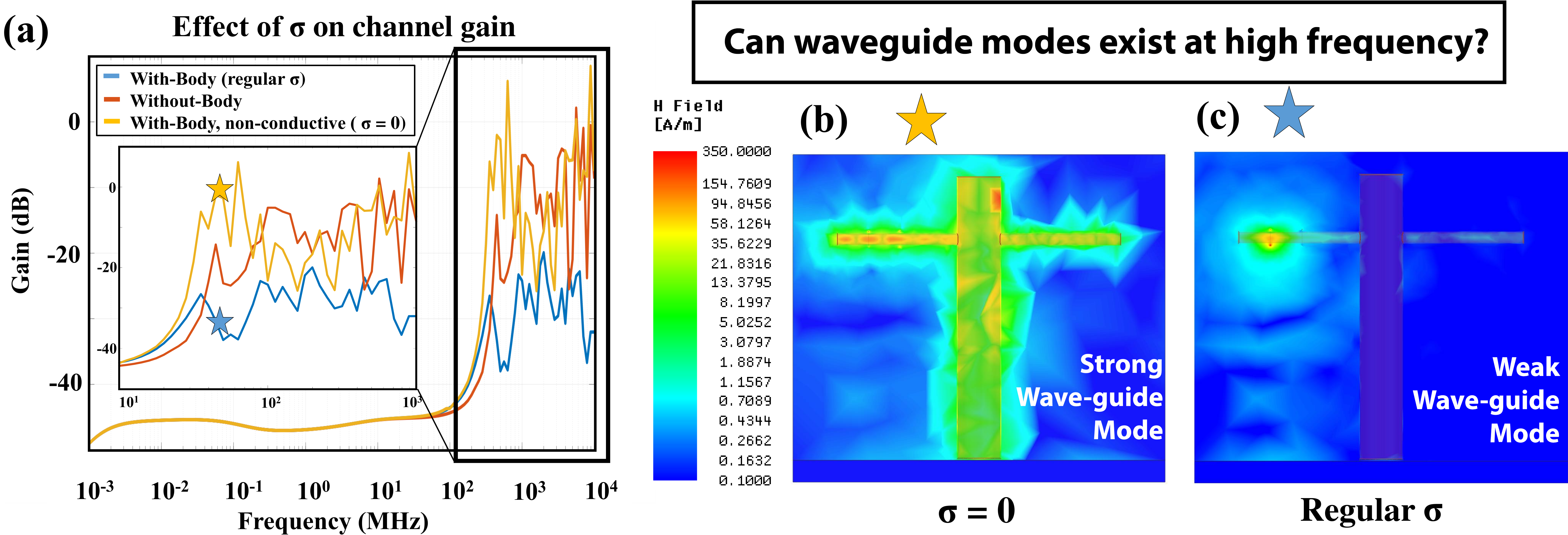}
    \caption{(a) Examining the change in channel gain, when $\sigma$ is artificially set to zero. Overall response becomes higher for f$>$100 MHz compared to the regular $\sigma$ case, implying that body attenuates Magnetic HBC fields at higher frequency. (b) H field plot for the $\sigma$ = 0 model at 447MHz shows Strong waveguide mode. (c) For the regular $\sigma$ model, the waveguide mode does exist at the same frequency, but is extremely weak.
}
    \label{fig:sigma0_plot}
\end{figure*}

Of course, when the material properties of skin and muscle are updated with the accurate Gabriel dataset, the waveguide modes in the region 100 kHz - 10 MHz disappear, seconded by the H-field distribution plot shown in Fig. \ref{fig:waveguide_low_freq}(e). At this point, we note that according to Gabriel dataset, the $\lambda$ inside human body becomes comparable to huuman body dimensions ($\sim$1m) for $f\sim100$ MHz, as can be seen in Fig. \ref{fig:waveguide_low_freq}(d). So technically, the human body \textit{should} be able to support waveguide modes for $f>$ 100 MHz. In reality however, as we saw before, the presence of the body in fact reduces signal transfer in this frequency region. Clearly, the effect cannot be explained solely by $\epsilon_r$ and this brings us to the final piece of the puzzle - the conductivity of the human body tissues.

\subsection{Effect of Conductivity ($\sigma$)}

% \[\nabla \times E = -\frac{dB}{dt}\]
% \[I = q \sigma AE\]

We recall that human body tissues are mildly conductive, and the conductivity increases with frequency, as shown in Fig. \ref{fig:sim_setup}(b). Whenever there is an alternating magnetic field present in or around a conductive object, part of the energy stored in the field should get absorbed by the object, as the alternating magnetic field would give rise to Eddy currents in the material, which would in-turn dissipate energy in the form of Ohmic loss in the conductive material. To examine whether this is actually the reason human body attenuates signal transfer in the high frequency ($f>100$ MHz), we perform a simulation with artificially setting the conductivity ($\sigma$) of skin and muscle to zero - as that would eliminate absorption due to conductivity.

The resulting transfer characteristics, given in Fig. \ref{fig:sigma0_plot}(a), shows that the non-conductive body model \textit{indeed} has overall higher response compared to the regular conductive model in the high frequency. Further, the response for the non-conductive model also becomes higher compared to the without-body model in the 100 MHz - 1 GHz frequency range. The response in this region shows waveguide like peaks, e.g. at f $=$ 447 MHz, where $\lambda_{muscle} = 8.9cm \approx$ diameter of the arm. The H-Field distribution plot for f$=$447 MHz, zero $\sigma$ case, given in Fig. \ref{fig:sigma0_plot}(b), shows a strong waveguide mode. Coming back to the conductive model, when we plot the H-Field distribution for the same frequency at regular $\sigma$, we see that indeed there is a very weak, albeit visible waveguide mode (Fig. \ref{fig:sigma0_plot}(c)). 

So in conclusion, even if the human body can support waveguide modes for $f>100$ MHz, these modes are severely attenuated due to the conductive nature of the human body leading to Eddy current loss (Fig. \ref{fig:eddy_current_plot}), and as a result, presence of the human body reduces the overall signal transfer in this frequency range, compared to the without-body case. 

\begin{figure}
    \centering
    \includegraphics[width=\columnwidth]{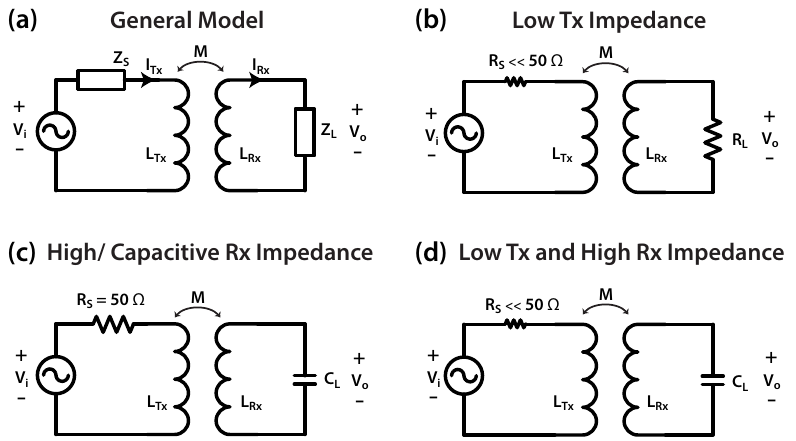}
    \caption{Simplified circuit model for MQS HBC coupling, for (a) General model, (b) Low transmitter impedance, (c) High/Capacitive receiver impedance, and (d) Both low transmitter impedance and high/capacitive receiver impedance.}
    \label{fig:mqs_ckt}
\end{figure}

\begin{figure*}
    \centering
    \includegraphics[width=\textwidth]{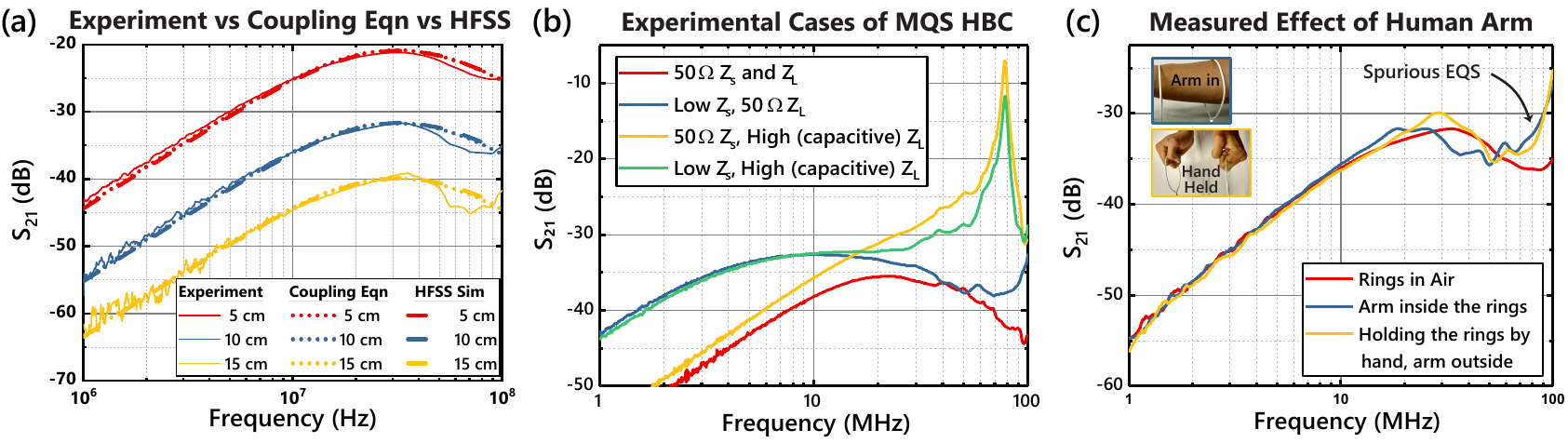}
    \caption{(a) Comparison between different methods of calculating $S_{21}$ for VNA measurements for various distances between axially aligned Tx and Rx coil. The results from experimental measurement, HFSS simulation for an identical setup, and MQS HBC coupling equation i.e. eqn. (\ref{eqn:general_equation}) all match with each other well. (b) Different Experimental MQS HBC modes, based on Tx and Rx impedances (c) MQS HBC gain vs frequency, comparing rings-in-air with arm-in-ring cases, demonstrating minimal effect of the human body channel in MQS HBC. An additional measurement with arm-outside, hands holding the rings is performed to detect spurious EQS coupling.}
    \label{fig:expt_results}
\end{figure*}

\section{Practical Excitation Model for Magnetic HBC}
\label{sec:expt}

As we had already mentioned in Sec. \ref{sec:excitation_model}, the excitation model used so far for EM simulations is an ideal current loop based model. In a practical application scenario, one needs to consider the source and load impedance present at the transmitter and the receiver respectively, as presented by the lumped circuit model in Fig. \ref{fig:mqs_ckt}(a). Also, we have seen in the previous section that there is a clear distinction between \textit{Magneto-Quasistatic (MQS) region}, and \textit{EM region} of the transfer characteristics. As shown in Fig. \ref{fig:waveguide_low_freq}(c), the wavelength of EM waves inside human body starts becoming comparable to the body dimensions, only for frequencies greater than 30 MHz. In fact for the ring-on-arm excitation case, EM effects start dominating only beyond 100 MHz and we see that in the EM region, the presence of human body hurts and does not help the case of Magnetic HBC. So for practical application scenarios, we are going to consider lower-frequency MQS domain - where quasistatic approximation\cite{das_nsr_2019} can be used, and the transmission channel can be modelled as a simple electrical circuit. This in turn justifies the lumped circuit model of Fig \ref{fig:mqs_ckt}(a).

\subsection{Magneto-quasistatic (MQS) HBC: Coupling Equation} 
% For MQS HBC, the frequency of operation is low enough (f$<$100 MHz in this case) so that the wavelength is large compared to the dimensions of the system. In that case, quasistatic approximation can be used \cite{das_nsr_2019}, and the transmission channel can be modeled as a simple electrical circuit. 
As shown in Fig. \ref{fig:mqs_ckt}(a), the general circuit model of MQS HBC consists of two inductors $L_{Tx}$ and $L_{Rx}$, representing the transmitter and the receiver rings respectively. A mutual inductance $M$ exists between the two inductors - this determines the relation between currents $I_{Tx}$ and $I_{Rx}$. Assuming a source impedance $Z_S$ and a load impedance $Z_L$, the following system of coupled differential equations represent the complete system:
\begin{equation}
\widetilde{V}_i-\widetilde{I}_{Tx}Z_S+M\dv{\widetilde{I}_{Rx}}{t}-L_{Tx}\dv{\widetilde{I}_{Tx}}{t}=0    
\label{eqn:cpldiff_tx}
\end{equation}
\begin{equation}
M\dv{\widetilde{I}_{Tx}}{t}-L_{Rx}\dv{\widetilde{I}_{Rx}}{t}-\widetilde{I}_{Rx}Z_L=0  
\label{eqn:cpldiff_rx}
\end{equation}

Given the source voltage is harmonic i.e. $\widetilde{V_i}=V_ie^{j\omega t}$, harmonic solutions of $I_{Tx}$ and $I_{Rx}$ can be obtained from eqn. (\ref{eqn:cpldiff_tx}) and (\ref{eqn:cpldiff_rx}) as:

\begin{equation}
    \widetilde{I}_{Tx}=\widetilde{V}_i/\left(Z_S+\frac{\omega^2 M^2}{j\omega L_{Rx}+Z_L}+j\omega L_{Tx}\right)
    \label{eqn:tx_curr}
\end{equation}

\begin{equation}
    \widetilde{I}_{Rx}=\frac{j\omega M\widetilde{I}_{Tx}}{j\omega L_{Rx}+Z_L}
    \label{eqn:rx_curr}
\end{equation}

The output voltage $V_o$ measured across the load impedance $Z_L$ can also be derived, giving the voltage gain of the system as:

\begin{equation}
    \frac{\widetilde{V}_o}{\widetilde{V}_i}=\frac{j\omega MZ_L}{(j\omega L_{TX}+Z_S)(j\omega L_{RX}+Z_L)+\omega^2M^2}
    \label{eqn:general_equation}
\end{equation}

Eqn (\ref{eqn:general_equation}) represents the voltage coupling/transfer for a general MQS HBC Tx-Rx pair. In the following subsection, we will go through a few special cases of source and load impedances.

\subsection{Termination Choices: Input and Output Impedance Variations and Results}

For a regular off-the-shelf RF device such as a Vector Network Analyzer (VNA), the common source and load impedance would be $Z_0=50\Omega$, and this in-fact is our default analysis and measurement mode. It had been shown however for Electroquasistatic Capacitive HBC \cite{maity_biophysical_2018}, that a low source impedance and a high/capacitive load impedance provides the best signal transfer - not a $50\Omega$-$50\Omega$ configuration. Keeping that in mind, we examine the following four termination cases (fig. \ref{fig:expt_results}b) for MQS HBC:

\subsubsection{VNA: $Z_{S}=Z_{L}=50\Omega$} 
Assuming the transmitting and receiving rings to be identical such that $L=L_{TX}=L_{RX}$ and $Z_0=Z_{S}=Z_{L}$, eqn. (\ref{eqn:general_equation}) reduces to:

\begin{equation}
    \frac{\widetilde{V}_o}{\widetilde{V}_i}=\frac{j\omega MZ_0}{(j\omega L+Z_0)^2+\omega^2M^2}
    \label{eqn:50ohm_equation}
\end{equation}

Further, assuming $M$ to be small with respect to $L$, the frequency for peak voltage gain or least loss in eqn. (\ref{eqn:50ohm_equation}) can be shown to be:
\begin{equation}
f_{peak} = Z_0/(2 \pi L)
\label{eqn:general_peak_freq}
\end{equation}

Fig. \ref{fig:expt_results}(a) shows this frequency behavior, comparing results from experiments, EM simulation of an identical setup in HFSS, and eqn. (\ref{eqn:50ohm_equation}). The details of the experiment setup has been given in the following section. As evident from fig. \ref{fig:expt_results}(a), results for the three cases line up with each other pretty well. The characteristics show a peak at around 30 MHz, with 20 dB/decade high pass and low pass slopes on both sides. Further, the gain is proportional to the mutual inductance $M$ and that in turn follows a inverse power law with respect to distance, so the gain falls as the distance is increased between the two coils.

% Additionally, as shown in Fig. \ref{fig:mqs_ckt}, there can be special operation cases depending on source and load impedances, discussed in detail in the following subsection.

% \subsubsection{MQS with Low Input Resistance}
% \begin{equation}
%     \frac{100j\omega M}{j\omega L (50 + j\omega L) + \omega^2 M^2}
% \end{equation}

% \subsubsection{MQS with High Load Resistance}
% \begin{equation}
%     \frac{2j\omega M}{j\omega L + 50}
% \end{equation}

% \subsubsection{MQS with Low Input Resistance and High Load Resistance}
% \begin{equation}
%     \frac{2M}{L}
% \end{equation}

% In the results subsection, we explore these different cases through transfer characteristics measurement experiments.

% \begin{figure}
%     \centering
%     \includegraphics[width=0.95\columnwidth]{comparison_plot.pdf}
%     \caption{Comparison between different methods of calculating $S_{21}$ for VNA measurements for various distances between axially aligned Tx and Rx coil. The results from experimental measurement, HFSS simulation for an identical setup, and MQS HBC coupling equation i.e. eqn. (\ref{eqn:general_equation}) all match with each other well. }
%     \label{fig:comp_plots}
% \end{figure}

\subsubsection{Low Input Impedance}
When source impedance is small, i.e. $Z_s \approx 0$, eqn. (\ref{eqn:general_equation}) reduces to
\begin{equation}
    \frac{V_o}{V_i}=\frac{j\omega MZ_L}{j\omega L_{TX}(j\omega L_{RX}+R_L)+\omega^2M^2}
     \label{eqn:low_input_equation}
\end{equation}
With a reduction in source impedance, the high pass slope moves toward the left - as can be seen by comparing the red and blue curves in fig. \ref{fig:expt_results}(b). In the ideal limit of zero input impedance, the high pass slope would completely disappear, making the low frequency characteristics completely flat.

\subsubsection{High RX (Capacitive)} With an increasing load resistance, the high frequency low pass slope would start moving towards right, vanishing in the limit of infinite load resistance or open circuit. In practice however, a finite load capacitance $C_L$ would be present due to parasitic coupling even for the open circuit scenario. In that case, eqn. (\ref{eqn:general_equation}) reduces to:
\begin{equation}
    \frac{V_o}{V_i}=\frac{j\omega M}{(1-\omega L_{RX}C_L)(j\omega L_{TX}+R_S)+j\omega^3C_LM^2}
    \label{eqn:high_load_equation}
\end{equation}

The load capacitance along with the inductance of the receiver coil would create a sharp resonance; with $C_L$ and $L_{TX}$ cancelling at $f = 1/(2\pi  L_{RX}C_L)$. The measurement result for this case is the yellow curve in fig. \ref{fig:expt_results}(b). showing a sharp peak at $\sim80$MHz.

\subsubsection{Low Input Resistance and High Load Resistance}
Finally, combining the two previous cases, we have:
\begin{equation}
    \frac{V_o}{V_i}=\frac{j\omega M}{j\omega L_{TX}(1-\omega L_{RX}C_L)+j\omega^3C_LM^2}
    \label{eqn:low_input_high_load_equation}
\end{equation}
This would show both a low frequency flat region (or, in non-ideal conditions, a high pass slope that has moved left with respect to 50$\Omega$ termination) and a high frequency sharp resonance peak. This behavior is seen in the measurement results, shown here as the green curve in fig. \ref{fig:expt_results}(b).

\subsection{Measurement Methodology}
\begin{figure}
    \centering
    \includegraphics[width=\columnwidth]{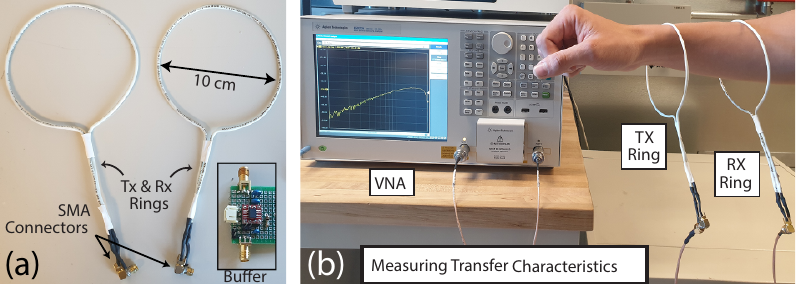}
    \caption{(a) The rings used as Tx and Rx for experiments in MQS HBC. The Buffer shown in the inset is used to provide low Tx impedance and/or high Rx impedance (b) Experimental setup to measure signal transfer characteristics using a VNA.}
    \label{fig:expt_setup}
\end{figure}

To perform measurements and compare different termination cases mentioned above, rings of diameter 10 cm are built out of 14 AWG insulated wire, as shown in Fig. \ref{fig:expt_setup}(a). Inductance of the rings are measured to be about 260 nH using a LCR meter. For experiments that require the presence of the human body, the rings are worn on the arm of a human subject as shown in fig. \ref{fig:expt_setup}(b). The experiments involving human subjects were approved by the Purdue University Institutional Review Board (IRB protocol \#1610018370). To perform measurements for the 50$\Omega$ termination case, the rings are directly connected to the ports of a Vector Network Analyzer through co-ax cables. Additionally a buffer setup is created (Fig. \ref{fig:expt_setup}(b)) using BUF602ID from Texas Instruments, to provide a high load impedance ($Z_L$) to the receiver ring, and/or a low source impedance ($Z_S$) to the transmitter ring. Placing the buffer between the VNA transmit port and the Tx coil provides a low source impedance, whereas placing it between the Rx coil and VNA receive port provides a high load impedance. These two cases can be combined by using buffers both at Tx and Rx. Measured $S_{21}$ for these different cases with the rings suspended in air at a distance of 10 cm is shown in Fig. \ref{fig:expt_results}(b). As expected from the discussion in the previous subsection, $S_{21}$ for the 50$\Omega$ termination case peaks at $f=50/(2*\pi*260)$ Hz or 31 MHz, as per Eqn. (\ref{eqn:general_peak_freq}). For low $Z_S$ the 20 dB/decade high-pass slope moves further left, and for high $Z_L$ the low-pass slope moves right. As discussed in the previous section, we see sharp resonance peaks for high $Z_L$ cases, where the capacitive component of the buffer input impedance cancels the inductance of the receiver ring. Note that in all these measured results, the \textit{TX and RX are axially aligned, and close to each other}. The channel gain would be lower when they are not aligned, or further from each other.

\begin{figure*}
    \centering
    \includegraphics[width=0.9\textwidth]{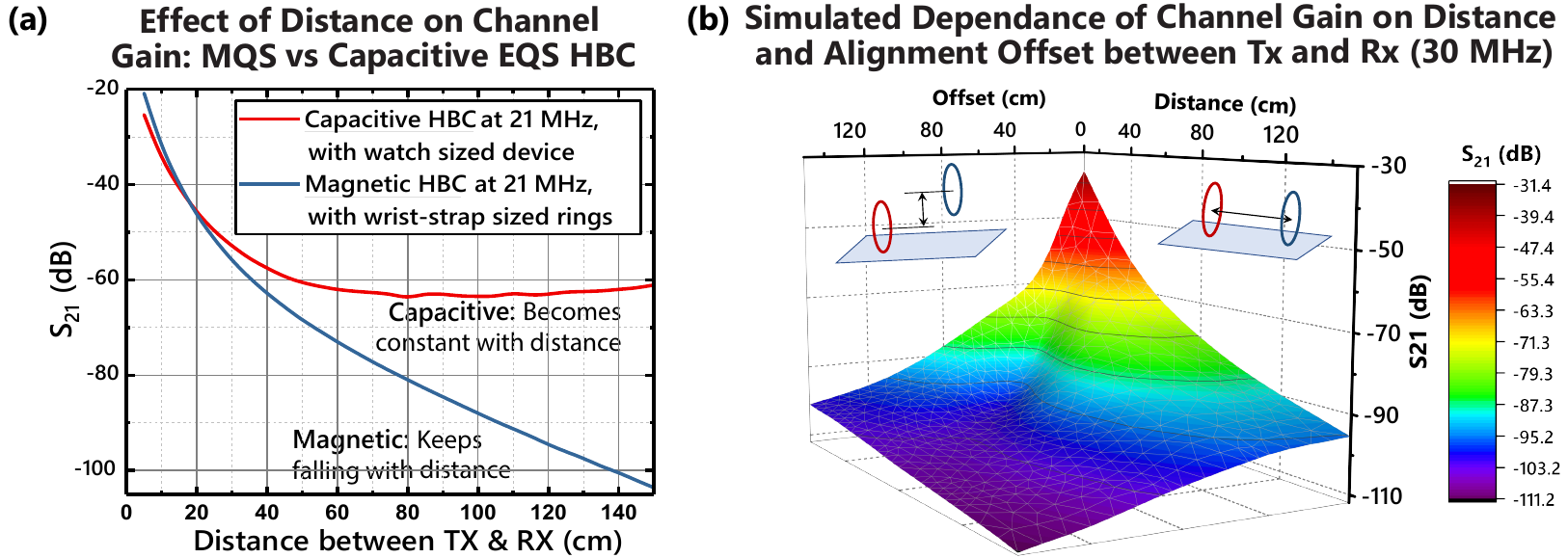}
    \caption{(a)  Comparison of Capacitive EQS HBC with MQS HBC channel gain, over distance. (b) HFSS simulation results showing sharp decline in channel gain with increase in distance and/or offset in alignment between Tx and Rx rings. Simulations are performed replicating our experimental setup, at it's peak frequency of 30 MHz (As shown before in Fig. \ref{fig:expt_results}(a)).}
    \label{fig:dist_off_sweep}
\end{figure*} 

\subsection{Effect of Human arm and Spurious EQS coupling}
Fig. \ref{fig:expt_results}(c) shows measured effects of human arm on MQS signal transfer for 50$\Omega$ termination at transmitter and receiver. For the most part, the results for the ``in-air" and ``arm-in" align with each other, deviating slightly in the 10 MHz- 100 MHz region, showing minimal effect of the presence of the human arm in the results. At this point, we note that the rings used as TX and RX can potentially also function as EQS electrodes, by capacitively coupling with the human body - this would basically be a case of a weak capacitive EQS HBC; and the gain in the signal transfer caused by this can be confused with MQS HBC gain provided by the human body. To eliminate this confusion, we also collect signal-transfer data with the human subject holding the two rings by hand - in this case, the human body cannot contribute to MQS HBC as the arm is not present inside the rings anymore, but spurious EQS coupling can still happen. This method confirms that the increased signal transfer in Fig. \ref{fig:expt_results}(c) near 100 MHz is indeed caused by spurious EQS coupling, and not by MQS HBC. So as we had seen in the simulations, \textit{the human body indeed has a minimal role in aiding signal transfer in MQS HBC}. Note that depending on experimental setup and Tx/Rx designs, the effect of spurious EQS coupling may become more pronounced, and may lead to less accurate results that might show the human body efficiently transferring MQS HBC signals. This could be one of the reasons of the conclusions drawn by \cite{park_embc_2015} regarding magnetic HBC body channel.

\subsection{Effect of Distance and Alignment: Comparison with Capacitive EQS-HBC for Wearable Applications}
In the final sets of results, we draw a comparison between Capacitive EQS and MQS HBC signal transfer, in the context of distance between the transmitter and receiver, as well as their placement or alignment on a user's body. Firstly to show the effect of distance, a watch sized capacitive HBC device is assumed - to keep the device size similar to the MQS case - where the rings can be assumed to be part of wrist-straps of watches. For Capacitive HBC, as shown in Fig. \ref{fig:dist_off_sweep}(a), the response first drops with distance when the TX and RX devices are very close, but becomes constant beyond a certain distance ($\sim$50 cm). For the MQS HBC case on the other hand, the response keeps dropping with distance. In fact in this typical example, MQS HBC is only stronger compared to EQS HBC, when the TX and RX are in extreme close proximity ($<$10 cm). In general, works by Das et al.\cite{das_nsr_2019} and Maity et al.\cite{maity_biophysical_2018} have shown that the gain in EQS capacitive becomes constant beyond a certain distance on the human body - irrespective of the specific positions of Tx and Rx on the body. For MQS however - since the body does not help in signal transfer - the gain keeps decreasing with increasing distance between Tx and Rx.

Further, to ensure good signal transfer, Tx and Rx in MQS HBC need to be well aligned. All of our previous results have assumed axial alignment between Tx and Rx. The reduction in gain when the alignment is off, is demonstrated in Fig. \ref{fig:dist_off_sweep}(b). The surface contour plot of Fig. \ref{fig:dist_off_sweep}(b) demonstrates the fact that MQS HBC can provide good signal transfer as long as Tx and Rx are placed close to each other, and are well aligned. Increase in either distance or offset results into a sharp decline in channel gain. This leads to the conclusion that for general wearable HBC application scenarios - only when low distance and axial alignment between Tx and Rx are possible MQS HBC could be useful. For a variety of other cases where such alignment is not possible EQS based devices are more suitable over similarly sized MQS based devices.

\section{Conclusion}
In conclusion, we develop the first theoretical modelling of Magnetic and MQS HBC, and show that the human body channel cannot fundamentally contribute to Magnetic Human Body Communication Channel Gain. In the low-frequency Magneto-Quasistatic region, the body does not provide any additional gain; and in the high-frequency Electromagnetic region, while it is possible for the body to support waveguide modes, those modes are highly attenuated by the conductive skin and muscle tissues of the body. We demonstrate different cases of MQS coupling through experiments involving varied transmitter and receiver impedance. Finally, we show that while at longer distances Capacitive EQS HBC has higher gain compared to MQS HBC, it could be applicable for uses cases where the transmitter and the receiver are aligned, and sufficiently close to each other.

\section*{Acknowledgements}
This work was supported in part by Air Force Office of Scientific Research YIP Award under Grant FA9550-17-1-0450, the National Science Foundation CRII Award under Grant CNS 1657455 and Eli Lilly and Company through the Connected Health care initiative.

\bibliographystyle{./bibliography/IEEEtran}
\bibliography{./bibliography/IEEEabrv,./bibliography/DATE_2020}

\end{document}